\documentclass[fleqn,10pt]{wlscirep}
\usepackage[utf8]{inputenc}
\usepackage[T1]{fontenc}
\usepackage[version=4]{mhchem}
\title{A Self-supervised Multimodal Deep Learning Approach to Differentiate Post-radiotherapy Progression from Pseudoprogression in Glioblastoma}

\usepackage{graphicx}
\usepackage{subfigure}
\usepackage{xcolor}
\usepackage{multirow}
\usepackage{bbm}
\usepackage{cleveref}
\usepackage{svg}

\usepackage{booktabs, multirow} % for borders and merged ranges
\usepackage{soul}% for underlines
\usepackage{xcolor,colortbl} % for cell colors
\usepackage{changepage,threeparttable} % for wide tables
\usepackage{booktabs} % For better table design

\author[1,2,3,*]{Ahmed Gomaa}
\author[1,2,3]{Yixing Huang}
\author[1,2,3]{Pluvio Stephan}
\author[4]{Katharina Breininger}
\author[1,2]{Benjamin Frey}
\author[2,5]{Arnd Dörfler}
\author[2,7]{Oliver Schnell}
\author[2,7]{Daniel Delev}

\author[7]{Roland Coras}
\author[1,2,3]{Charlotte Schmitter}
\author[8]{Jenny Stritzelberger}
\author[1,2,3]{Sabine Semrau}
\author[9]{Andreas Maier}
\author[9]{Siming Bayer}
\author[3, 10]{Stephan Schönecker}

\author[6,7, 11]{Dieter H Heiland}
\author[12,13]{Peter Hau}
\author[1,2,3]{Udo S. Gaipl}
\author[1,2,3]{Christoph Bert}
\author[1,2,3]{Rainer Fietkau}
\author[2,3,5,**]{Manuel A. Schmidt}
\author[1,2,3,**]{Florian Putz}
\affil[1]{Department of Radiation Oncology, University Hospital Erlangen, Friedrich-Alexander-Universität Erlangen-Nürnberg, Erlangen, 91054, Germany}

\affil[2]{Comprehensive Cancer Center Erlangen-EMN (CCC ER-EMN), Erlangen, 91054, Germany}

\affil[3]{Bavarian Cancer Research Center (BZKF), Erlangen, 91052, Germany.}

\affil[4]{Institute of Neuroradiology, University Hospital Erlangen, Friedrich-Alexander-Universität Erlangen-Nürnberg, Erlangen, 91054, Germnay}

\affil[5]{Universität Würzburg, Center for Artificial Intelligence and Data Science, Würzburg, 97074, Germany}

\affil[6]{Translational Neurosurgery, Alexander-Friedrich-Universität Erlangen-Nürnberg, Erlangen, 91054 Germany}

\affil[7]{Department of Neurosurgery, University Hospital Erlangen, Friedrich-Alexander-Universität Erlangen-Nürnberg, Erlangen, 91054, Germany}

\affil[8]{Department of Neurology, University Hospital Erlangen, Friedrich-Alexander-Universität Erlangen-Nürnberg, Erlangen, 91054, Germany}

\affil[9]{Pattern Recognition Lab, Friedrich-Alexander-Universität Erlangen-Nürnberg, Erlangen, Germany}

\affil[10]{Department of Radiation Oncology, University Hospital Ludwig Maximilian University of Munich, 81377 Munich, Germany}

\affil[11]{Department of Neurological Surgery, Northwestern University Feinberg School of Medicine, Chicago, IL, 60611 USA}

\affil[12]{Department of Neurology, University Hospital Regensburg, Regensburg, Germany}
\affil[13]{Wilhelm Sander-NeuroOncology Unit, University Hospital Regensburg, Regensburg, Germany}

\affil[*]{ahmed.gomaa@uk-erlangen.de}
\affil[**]{Contributed equally}

\begin{abstract}
Accurate differentiation of pseudoprogression (PsP) from True Progression (TP) following radiotherapy (RT) in glioblastoma patients is crucial for optimal treatment planning. However, this task remains challenging due to the overlapping imaging characteristics of PsP and TP. 
This study therefore proposes a multimodal deep-learning approach utilizing complementary information from routine anatomical MR images, clinical parameters, and RT treatment planning information for improved predictive accuracy. 
The approach utilizes a self-supervised Vision Transformer (ViT) to encode multi-sequence MR brain volumes to effectively capture both global and local context from the high dimensional input. 
The encoder is trained in a self-supervised upstream task on unlabeled glioma MRI datasets from the open BraTS2021, UPenn-GBM, and UCSF-PDGM datasets (n = 2317 MRI studies) to generate compact, clinically relevant representations from FLAIR and T1 post-contrast sequences. 
These encoded MR inputs are then integrated with clinical data and RT treatment planning information through guided cross-modal attention, improving progression classification accuracy.
This work was developed using two datasets from different centers: the Burdenko Glioblastoma Progression Dataset (n = 59) for training and validation, and the GlioCMV progression dataset from the University Hospital Erlangen (UKER) (n = 20) for testing. The proposed method achieved competitive performance, with an AUC of 75.3\%, outperforming the current state-of-the-art data-driven approaches.
Importantly, the proposed approach relies solely on readily available anatomical MRI sequences, clinical data, and RT treatment planning information, enhancing its clinical feasibility. 
The proposed approach addresses the challenge of limited data availability for PsP and TP differentiation and could allow for improved clinical decision-making and optimized treatment plans for glioblastoma patients.

\end{abstract}
\begin{document}

\flushbottom
\maketitle
\thispagestyle{empty}

% \noindent Please note: Abbreviations should be introduced at the first mention in the main text – no abbreviations lists. Suggested structure of main text (not enforced) is provided below.

\section*{Introduction}

Glioblastoma is the most frequent primary brain tumor among adults \cite{price2024cbtrus}, and understanding its treatment prognosis is critical for optimal treatment selection and patient management  \cite{ostrom2020cbtrus, hagag2024deep, zhou2024integrated, Lau2014Molecularly}. Despite advancements in surgical techniques, radiotherapy (RT), and chemotherapy, the prognosis for glioblastoma patients remains poor, with median survival times ranging from $12$ to $15$ months \cite{stupp2005radiotherapy, koshy2012improved}. A particularly immense challenge in the management of glioblastoma is distinguishing between pseudoprogression (PsP) and true progression (TP) following chemoradiation therapy. PsP refers to RT-related transient worsening of radiographic images that mimics tumor growth but does not indicate actual tumor progression and usually resolves spontaneously. TP, in contrast, refers to real tumor growth and proliferation, necessitating a change in a patient's treatment. The treatment options for TP, including surgery and re-irradiation, carry a significant risk of toxicity. Moreover, PsP being mainly induced by RT-related vascular damage, inflammation, and necrosis, can significantly worsen with a second course of radiation \cite{brandsma2009pseudoprogression}. 

Misinterpretations can lead to premature discontinuation of effective treatments in the case of PsP or delayed intervention in the case of TP, potentially worsening patient outcomes. Therefore, the precise differentiation between PsP and TP is crucial for optimal clinical decision-making. While surgical biopsy serves as a reliable approach for early tumor progression diagnosis, it is not without limitations. The invasive nature of tissue biopsy can represent a major concern, limiting repeated procedures. Moreover, this concern is further compounded by potential inaccuracies resulting from biopsy site selection, and mixed histological patterns can considerably limit its diagnostic accuracy \cite{da2011pseudoprogression, yang2020adding}. Furthermore, a biopsy may not be feasible for certain anatomical regions with post-treatment imaging changes \cite{da2011pseudoprogression}. These limitations highlight the need for complementary, non-invasive predictive tools with improved accuracy and reliability.

With the rapid advancement of artificial intelligence, there has been a growing interest in applying machine learning and deep learning methods to improve diagnostic accuracy and treatment planning using medical data \cite{huang2024principles, erdur2024deep}. For instance, H. Akbari et al. employed a pre-trained Convolutional Neural Network (CNN) to automatically extract features from multiparametric MRI scans. 
These features were subsequently concatenated with extracted radiomics features to classify progression status using Support Vector Machines (SVM). This approach harnesses the power of deep learning to extract and analyze complex features from imaging data, showing promising results with higher accuracy rates compared to manual interpretations \cite{akbari2020histopathology}.
Using multiparametric MRI data, Lee et al. proposed a hybrid deep learning architecture for differentiating high-grade glioma (HGG) tumor progression from treatment-related changes. 
Their proposed model consisted of a CNN integrated with Long Short-Term Memory (LSTM) units and was trained on a dataset of $43$ biopsy-proven HGG patients. 
Five standard MRI sequences (DWI, T2, FLAIR, T1 pre- and post-contrast) were combined with two calculated sequences (T1 post-contrast $-$ T1 and T2 $-$ FLAIR) to provide additional information for model training. The authors demonstrated that incorporating all available modalities led to the best performance, achieving a mean area under the receiver operating characteristic curve (AUC) of $0.81$ in a three-fold cross-validation setup. However, this evaluation did not include an external test set \cite{lee2020discriminating}. 
Moreover, Moassefi et al. leveraged transfer learning by utilizing a pre-trained 3D-DenseNet-121 architecture, originally developed for the sequence registration task. This pre-trained model was subsequently finetuned on a dataset of $124$ cases in a five-fold cross-validation. Their approach achieved an AUC of $0.88$, though their evaluation was also not conducted on an external test set \cite{moassefi2022deep}.

Other works have investigated a multimodal approach that leverages the diagnostic information existing in the clinical data as well as the imaging data. 
To this end, Sun et al. developed a random forest model to discriminate PsP from TP. Their model incorporated radiomics features extracted from T1 post-contrast MRI scans together with clinical information, such as patient demographics (sex, age), performance status (KPS score), extent of surgical resection, neurological deficits, and mean radiation dose \cite{sun2021differentiation}. 
Additionally,  Jang et al. proposed a CNN-LSTM hybrid architecture that was trained using $9$ transversal post-contrast T1 image slices, that were centered on the growing lesion, in combination with clinical information consisting of age, gender, molecular features, and RT dose and fractionation parameters \cite{jang2018prediction}. 

Although the current literature has shown promising advancements in the progression classification using imaging data, it has relied heavily on first-order Radiomics features and CNNs for processing medical images, which may be suboptimal for capturing long-range dependencies in high-dimensional MR inputs \cite{dosovitskiy2020image}. Additionally, the existing research on combining the imaging modality with clinical data is still limited, with the existing work relying on simple modality fusion techniques. This can lead to limited integration of complementary information across modalities, resulting in suboptimal feature representations and reduced accuracy \cite{wang2020makes, jang2018prediction}.

Therefore, this work proposes a novel deep learning approach combining a multimodal transformer-based architecture with a self-supervised MRI encoder. 
Unlike previous approaches that predominantly relied on CNNs, our methodology harnesses the attention mechanism capabilities of transformers to process high-dimensional MRI data more effectively. 
The attention mechanism of transformers enables them to capture both global and local context information within the data, making them well-suited for tasks involving complex spatial relationships \cite{he2023transformers, dosovitskiy2020image, raghu2021vision}.
Furthermore, the proposed work adopts a self-supervised learning paradigm, which offers several distinct advantages over traditional supervised approaches \cite{Chen2019Self, shurrab2022self}. 
Self-supervised learning allows the model to learn from a large amount of unlabeled data, which is readily available from various data sources.
This is particularly advantageous in the context of medical imaging data and progression classification, where labeled data is scarce and difficult to obtain. 
We hypothesize that by leveraging unlabeled data self-supervised learning could effectively exploit the rich information present in MRI scans, resulting in enhanced generalization and robustness in progression status classification \cite{shurrab2022self}. Additionally, the proposed model employs cross-modal attention to integrate patients' structural clinical features with corresponding imaging data, for enhanced predictive accuracy.

In summary, our work introduces three key contributions to the field of glioblastoma progression classification:
\begin{itemize}

\item A multimodal transformer-based architecture for differentiating PsP from TP is proposed and evaluated, demonstrating the advantage of the self- and cross-attention mechanism.

\item A self-supervised learning approach is adopted, benefiting from the large amounts of openly available unlabeled MRI data to overcome the data scarcity limitation and improve model generalizability.

\item We demonstrate that the proposed deep learning approach combining a multimodal transformer-based architecture with a self-supervised MRI encoder achieves improved predictive performance over the state-of-the-art models as well as simple transfer learning. 

\end{itemize}

To assess model performance and generalizability, we externally validate the proposed transformer architecture on an independent test set from a separate institution.

%%%%%%%%%%%%%%%%%% Methods %%%%%%%%%%%%%%%%%%

\section*{Methods}
% Topical subheadings are allowed. Authors must ensure that their Methods section includes adequate experimental and characterization data necessary for others in the field to reproduce their work.

\subsection*{Datasets}

\subsubsection*{Self-supervised Training} For the pre-text task, three public datasets for glioma cases were used (n = $2317$). These datasets encompass the University of California San Francisco Preoperative Diffuse glioma MRI (UCSF-PDGM) dataset, comprising $496$ cases \cite{calabrese2022university,clark2013cancer}, the Brain Tumor Segmentation (BraTS 2021) dataset with $1251$ cases \cite{baid2021rsnaasnrmiccai, menze2014multimodal, bakas2017advancing}, and the University of Pennsylvania Glioblastoma (UPenn-GBM) dataset, which includes $570$ cases \cite{bakas2022university, clark2013cancer}.

\subsubsection*{Progression Classification} For the training and validation of the proposed deep learning model, we employ the Burdenko's Glioblastoma Progression Dataset \cite{Zolotova2023Burdenko} (n = 180 patients). Patients were eligible for inclusion if they were diagnosed with either PsP or TP and had at least two morphologic MR series: a T1-weighted contrast-enhanced (T1CE) and a T2-Fluid-Attenuated Inversion Recovery (FLAIR) sequence, both acquired during the follow-up period when progression was identified. Out of the $180$ patients in the dataset, $59$ met these criteria. The MR volumes are supplemented with clinical information, including IDH mutation status, O[6]-methylguanine-DNA methyltransferase (MGMT) promoter methylation, age, gender, as well as the dates of RT and follow-imaging up scans. Each patient also has a corresponding RT study, which includes four morphologic MRI sequences (T1, T1CE, T2, and FLAIR), the planning CT scan, and the DICOM RT planning objects, including the three-dimensional dose distribution. Furthermore, to calculate the dose parameters for the enlarging tumor in the follow-up scans, we first segmented the tumors in the follow-up MR volumes using an nnU-Net, trained on the BraTS2021 dataset \cite{isensee2021nnu}. Subsequently, we rigidly registered the T1CE and FLAIR follow-up MR sequences at which the tumor enlargement was observed in the radiotherapy planning CT using Advanced Registration Tools (ANTs) version 0.5.3 \cite{avants2011reproducible}. Leveraging the calculated transform, we mapped the follow-up lesion segmentation into the frame-of-reference of the planning-CT and RT dose distribution to extract the dose-related features, namely mean, min, median, and D98 dose.

A second independent dataset from the prospective GlioCMV study (NCT02600065, GlioCMV UKER progression dataset, n = 20) acquired at the University Hospital Erlangen (UKER) was used as an external test set \cite{goerig2020early, goerig2016frequent}. This dataset includes similar clinical covariates and imaging modalities, allowing for external validation of the model’s performance in a real-world clinical setting. PsP and TP were differentiated in the GlioCMV UKER progression dataset either by histology (biopsy or tumor resection, n = 8) or by longitudinal imaging follow-up (n = 12). 
The ethics committee at the University Hospital Erlangen approved the conduction of the study (Approval Number: 265\_14 B) and all the patients had given their written informed consent for participation as well as for secondary scientific use of their data. Furthermore, all methods were conducted in accordance with the relevant guidelines and regulations.
The patient characteristics of each dataset are included in \Cref{tab:patChar}. In this supervised learning task, the model was trained using a 5-fold cross-validation approach on Burdenko's dataset, with folds stratified based on progression class. Each fold's trained model was then evaluated on the GlioCMV UKER dataset. Furthermore, the predictions of the test set are also ensembled using soft voting.

\begin{table}[h]
\centering
\begin{tabular}{lrr}
\hline
\textbf{Variable}                                                                                              & \textbf{Burdenko (n = 59)} & \textbf{GlioCMV UKER (n = 20)}   \\ \hline
\textbf{Gender}                                                                                                  &                 &               \\
\hspace{0.25cm} Female, n (\%)                                                                                          & 28 (47.5)              & 11 (55)            \\
\hspace{0.25cm} Male, n (\%)                                                                                            & 31 (52.5)             & 9 (45)             \\ \hline
\textbf{IDH Mutation Status }                                                                                    &                 &               \\
\hspace{0.25cm} Mutant, n (\%)                                                                                          & 8 (13.6)               & 1 (5)             \\
\hspace{0.25cm} Wildtype, n (\%)                                                                                        & 36 (61.0)              & 19 (95)            \\
\hspace{0.25cm} Unknown, n (\%)                                                                                         & 15 (25.4)              & 0 (0)             \\ \hline
\textbf{MGMT Methylation Status}                                                                                 &                 &               \\
\hspace{0.25cm} Methylated, n (\%)                                                                                      & 14 (23.7)              & 9 (45)            \\
\hspace{0.25cm} Unmethylated, n (\%)                                                                                    & 23 (40.0)             & 3  (15)           \\
\hspace{0.25cm} Unknown, n (\%)                                                                                         & 22 (37.3)              & 8 (40)            \\ \hline
% \begin{tabular}[c]{@{}l@{}}Median Time to Progression \\ Appearance (Weeks), median, range\end{tabular} &                 &               \\ \hline
\textbf{Progression Classification}                                                                              &                 &               \\
\hspace{0.25cm} True Progression, n (\%)                                                                                & 34 (57.6)             & 11 (55)           \\
\hspace{0.25cm} Pseudoprogression, n (\%)                                                                               & 25 (42.4)            & 9 (45)             \\ \hline
\textbf{Age in years}, median, range                                                                              & 57, (18-82)     & 58.5, (45-75) \\ \hline
\end{tabular}
\caption{Patient characteristics of the Burdenko’s Glioblastoma Progression Dataset (Burdenko) and the GlioCMV UKER progression dataset (UKER).}
\label{tab:patChar}
\end{table}

\subsection*{Data Preprocessing}
Each subject within the utilized datasets possessed a minimum of $2$ corresponding MRI modalities, namely T1 post-contrast and T2-FLAIR sequences. These modalities had already undergone preprocessing steps including spatial normalization, skull stripping, and bias field correction. The two morphological MRI scans underwent further processing before being passed into the deep learning model. This processing involved cropping each scan to a standardized size of $160 \times 160 \times 160$ voxels while maintaining an isotropic spacing of 1 mm. Subsequently, the cropped scans were concatenated along the channel dimension. Following this, histogram standardization and channel-wise z-normalization were applied to the data. 

For the structured clinical features, continuous values were normalized to have a mean of zero and a standard deviation of $1$. Categorical values, on the other hand, were encoded to one-hot encoded vectors. Moreover, features that were colinear with different features were dropped. 

To identify the optimum number of features, we performed SHapley Additive exPlanations analysis \cite{lundberg2017unified} on the cross-validation sets. After sorting the features based on their importance, as shown in \Cref{fig:SHAP}, we chose the first $M$ features leading to the highest AUC on the validation sets. For this setup, the optimum value of $M$ was $4$.

\subsection*{Model Architecture and Training Process}
\begin{figure}
    \centering
    \includegraphics[width=0.75\linewidth]{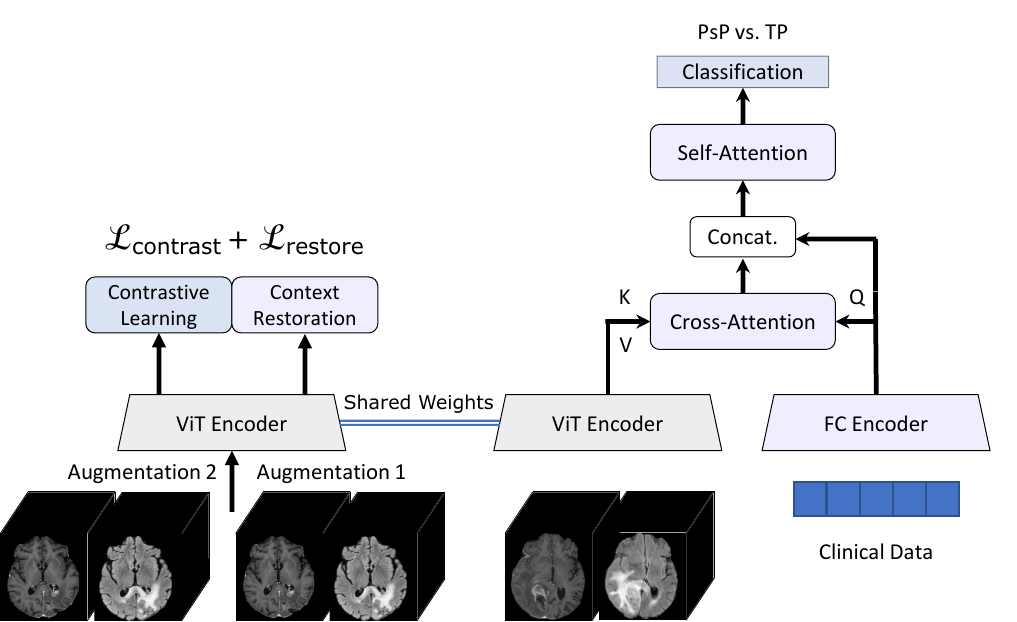}
    \caption{The proposed workflow of the model. On the left-hand side is the self-supervised setup. On the right-hand side is the progression classification setup after the self-supervised training. In the inference phase, multi-parameter MRI volumes and clinical data are fed to the pre-trained ViT encoder and FC encoder, respectively.}
    \label{fig:proposed-architecture}
\end{figure}

This work introduces a dual-phase deep learning architecture, as depicted in \Cref{fig:proposed-architecture}. In the first phase, a Vision Transformer (ViT) serves as an encoder for the MRI data. The parameters of this imaging encoder were optimized in a self-supervised manner using a large collection of unlabeled MRI scans utilizing two proxy tasks, namely context restoration and contrastive learning. This phase aims to employ the self-attention attention capabilities of the ViT encoder in acquiring discriminative and compact representation vectors from complex and high-dimensional MR datasets \cite{he2023transformers, azad2023advances}. Detailed information on the self-supervised training strategy is provided in the supplementary material.

In the second phase, the pre-trained self-supervised ViT with frozen parameters is used to encode the input MR volumes. The encoded volumes are represented as $E_{i} \in \mathbb{R}^{N \times d}$, where $N$ is the number of tokens extracted from the MR volume, and $d$ is the dimension on each token. The clinical features are encoded and reshaped into $E_{c} \in \mathbb{R}^{M \times d}$, where $M$ is the number of tokens derived from the reshaping process and ${d}$ is the shared dimension between the clinical and MR tokens. These encoded representations, $E_{i}$ and $E_{c}$, are then integrated via guided cross-attention, which captures the pair-wise interactions between the encoded modalities \cite{lu2019vilbert, gomaa2024comprehensive, tsai2019multimodal}. In this process, it is hypothesized that the encoded clinical data guides the cross-attention mechanism by focusing on the most relevant spatial regions in the MR volumes. Specifically, the guided cross-attention mechanism is formulated as follows:

\begin{equation}
\begin{aligned}
 E_{i \rightarrow c}  & = Cross\text-Attention_{i \rightarrow c}\left(E_i, E_c\right) \\ 
& =\operatorname{softmax}\left(\frac{Q_c K_i^{\top}}{\sqrt{d_k}}\right) V_i \\
& = \operatorname{softmax}\left(\frac{W_q E_c\left(W_k E_i\right)^{\top}}{\sqrt{d}}\right) W_v E_i, \\
\end{aligned}
\end{equation}

where $\{W_q, W_k, W_v\} \in \mathbb{R}^{d\times d}$ are learnable weight matrices. Subsequently, a self-attention layer is implemented to capture the intra-modal interactions within the vector resulting from the concatenation of $E_{i \rightarrow c}$ with $E_c$. The output of the self-attention module is then fed to two fully connected layers for output prediction after undergoing attention pooling. 

Given that in our implementation $M \ll N$, where $M=4$ and $N=512$, the encoded clinical vector guides the attention process and reduces the token length to $M$. This reduction in the token length considerably impacts the model's computational complexity, particularly when applying the self-attention mechanism in the following stage. Specifically, the complexity of the self-attention process is typically quadratic with respect to the input sequence length. Therefore, by reducing the sequence length from $N$ to $M$, the computation and memory cost is significantly reduced. 

\subsection*{Alternative Training Strategies}
We evaluated two commonly used training strategies to compare them with our proposed self-supervised approach. First, we assessed transfer learning, where a Vision Transformer (ViT) was trained on a survival prediction task using 592 samples from the UCSF-PDGM and UPenn-GBM datasets. Afterward, the trained ViT was used as an encoder for the MR data as part of the same multimodal architecture used in the proposed approach in the progression classification task. The second strategy involved training the entire multimodal architecture from scratch end-to-end for the progression classification task.

\section*{Results}

\subsection*{Comparison with the State-of-the-art}
\begin{figure}
    \centering
    \includegraphics[width=0.80\linewidth]{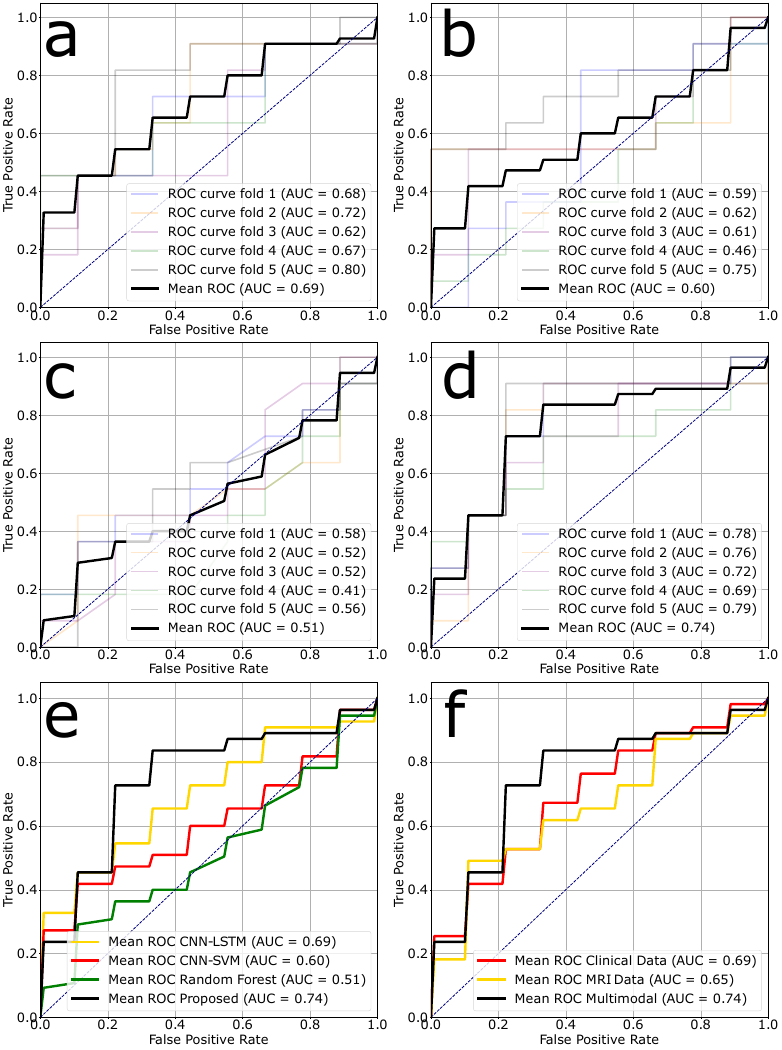}
    \caption{ROC curves evaluated on the external test set. Subplot (a) presents the ROC for the CNN-LSTM model (Jang et al. \cite{jang2018prediction}), while (b) displays the ROC for the CNN-SVM model (Akbari et al. \cite{akbari2020histopathology}). Subplot (c) illustrates the ROC for the random forest model (Sun et al. \cite{sun2021differentiation}), and (d) features the proposed model. Subplot (e) highlights the mean ROC curves of the proposed model and existing baselines, and (f) shows the mean ROC curves depending upon the input modality.}
    \label{fig:aucProposed}
\end{figure}

The proposed transformer-based architecture with a pre-trained imaging encoder achieved an AUC of 0.753 on the external test set. This was higher than the CNN-LSTM model by Jang et al. (0.686), the CNN-SVM model by Akbari et al. (0.677), and the Random Forest model by Sun et al. (0.530). In terms of accuracy, the proposed model achieved 0.750, compared to CNN-LSTM at 0.550, CNN-SVM at 0.650, and Random Forest at 0.550. For sensitivity, Random Forest had the highest value at 0.927, but with very low specificity of 0.111. The proposed model had a sensitivity of 0.727, which was higher than the CNN-LSTM model (0.455) and similar to CNN-SVM (0.7). Regarding specificity, the proposed model achieved 0.8, compared to CNN-LSTM at 0.625, CNN-SVM at 0.633, and Random Forest at 0.555. Overall, the model demonstrated an improved AUC compared to the state-of-the-art methods as well as a balanced sensitivity and specificity. These results are summarized in \Cref{tab:SOTA} and ROC curves are illustrated in \Cref{fig:aucProposed}(a-e).

\begin{table}[!htp]\centering
\scriptsize
\begin{tabular}{lcccccccccc}\toprule
\multirow{2}{*}{Model} &\multicolumn{4}{c}{Validation} & &\multicolumn{4}{c}{External Test} \\\cmidrule{2-5} \cmidrule{7-10}
&AUC &Accuracy &Sensitivity &Specificity & &AUC &Accuracy &Sensitivity &Specificity \\\midrule
 \multirow{2}{*}{CNN-LSTM: Jang et al. \cite{jang2018prediction}} &0.744 ± 0.095 &0.797 ± 0.104 &0.705 ± 0.165 &0.880 ± 0.036 & &0.698 ± 0.060 &0.590 ± 0.045 &0.491 ± 0.187 &0.711 ± 0.241 \\
         & -- & -- & -- & -- & & (0.686) & (0.550) & (0.455) & (0.625) \\
 \multirow{2}{*}{CNN-SVM: Akbari et al. \cite{akbari2020histopathology}} &0.527 ± 0.199 &0.558 ± 0.149 &0.619 ± 0.109 &0.480 ± 0.271 & &0.606 ± 0.092 &0.570 ± 0.020 &0.488 ± 0.093 &0.644 ± 0.113 \\
 & -- & -- & -- & -- & & (0.677) & (0.650) & (0.700) & (0.633) \\
 \multirow{2}{*}{Random Forest: Sun et al. \cite{sun2021differentiation}} &0.754 ± 0.110 &0.730 ± 0.094 &0.795 ± 0.144 &0.640 ± 0.233 & &0.516 ± 0.059 &0.560 ± 0.021 &0.927 ± 0.036 &0.111 ± 0.000 \\
& -- & -- & -- & -- & & (0.530) & (0.550) & (0.909) & (0.555) \\
 \multirow{2}{*}{Proposed} &0.883 ± 0.044 &0.770 ± 0.062 &0.819 ± 0.109 &0.720 ± 0.078 & &\textbf{0.748} ± 0.038 &\textbf{0.670} ± 0.090 &0.618 ± 0.118 &\textbf{0.733} ± 0.079 \\
 & -- & -- & -- & -- & & (0.753) & (0.750) & (0.727) & (0.800) \\
\bottomrule
\end{tabular}
\caption{Performance comparison of the proposed model with state-of-the-art models. Results after ensembling with soft voting are provided in parentheses.}
\label{tab:SOTA}
\end{table}

\subsection*{Unimodal vs. Multimodal}
The predictive performance, based on the input modality, was evaluated using AUC, accuracy, sensitivity, and specificity. The multimodal approach outperformed the others, achieving a mean AUC of $0.753$, accuracy of $0.750$, sensitivity of $0.727$, and specificity of $0.8$. In comparison, the model trained on structured clinical data alone, using a 4-layer fully connected network, reached a mean AUC of $0.727$, accuracy of $0.7$, sensitivity of $0.636$, and specificity of $0.778$. When using only imaging data, the finetuned self-supervised ViT model achieved a mean AUC of $0.717$, accuracy of $0.7$, sensitivity of $0.545$, and specificity of $0.857$. A detailed summary of these results is presented in \Cref{tab:resultsMain} and \Cref{fig:aucProposed}(f). %, and the corresponding AUC curves are visualized in \Cref{fig:aucProposed}.

\begin{table}[!htp]\centering
\begin{tabular}{lccccc}\toprule
Modality &AUC &Accuracy &Sensitivity &Specificity \\
\hline
\multirow{2}{*}{MRI} &0.659 ± 0.065 &0.620 ± 0.119 &0.454 ± 0.342 &0.822 ± 0.160 \\
& (0.717) & (0.700) & (0.545) & (0.857) \\
\multirow{2}{*}{Clinical Data} &0.698 ± 0.026 &0.630 ± 0.072 &0.618 ± 0.121 &0.644 ± 0.083 \\
& (0.727) & (0.700) & (0.636) & (0.778) \\
\multirow{2}{*}{Multimodal} &0.748 ± 0.038 &0.670 ± 0.090 &0.618 ± 0.118 &0.733 ± 0.079 \\
& (0.753) & (0.750) & (0.727) & (0.800) \\
\bottomrule
\end{tabular}
\caption{Performance on the external test based on input modality. Results after ensembling with soft voting are provided in parentheses.}
\label{tab:resultsMain}
\end{table}

\subsection*{Impact of Self-supervised Learning}
To study the effect of the ViT self-supervised pre-training on the predictive performance, we trained the model under three setups: End-to-end training from scratch, transfer learning, and self-supervised downstream training. For the end-to-end training, the model achieved an AUC of $0.697$, an accuracy of $0.6$, a sensitivity of $0.545$, and a specificity of $0.667$. In the transfer learning approach, the model was first pre-trained on survival prediction using $592$ MR samples and then fine-tuned for the progression classification task. The fine-tuned model achieved an AUC of $0.727$, an accuracy of $0.650$, a sensitivity of $0.636$, and a specificity of $0.7$. Lastly, in the self-supervised downstream training setup, the model achieved the highest overall performance, with an AUC of $0.753$, an accuracy of $0.75$, a sensitivity of $0.636$, and a specificity of $0.8$. These results, summarized in \Cref{tab:resultAblation}, indicate that self-supervised training led to better predictive performance, particularly in terms of AUC and sensitivity, compared to both end-to-end and transfer learning approaches.

\begin{table}[ht]
\centering
\begin{tabular}{lcccc}\toprule
Model & AUC &  Accuracy & Sensitivity &  Specificity\\
\hline
\multirow{2}{*}{ViT End-to-end} & 0.675 ± 0.037 & 0.600 ± 0.045 & 0.655 ± 0.134 & 0.533 ± 0.130 \\
& (0.697) & (0.600) & (0.545) & (0.667) \\
\multirow{2}{*}{ViT Transfer Learning} & 0.711 ± 0.060 & 0.612 ± 0.037 & 0.582 ± 0.136 & 0.645 ± 0.129 \\
& (0.727) & (0.650) & (0.636) & (0.700) \\
\multirow{2}{*}{ViT Self-supervised} &0.748 ± 0.038 &0.670 ± 0.090 &0.618 ± 0.118 &0.733 ± 0.079\\
& (0.753) & (0.750) & (0.636) & (0.800) \\
\bottomrule
\end{tabular}
\caption{Predictive performance of the multimodal model on the external test set based on the training method of the ViT imaging encoder, with the soft voting ensembling results reported in parentheses.}
\label{tab:resultAblation}
\end{table}

\subsection*{Feature Importance Analysis}
\begin{figure}
    \centering
    \includegraphics[width=0.65\linewidth]{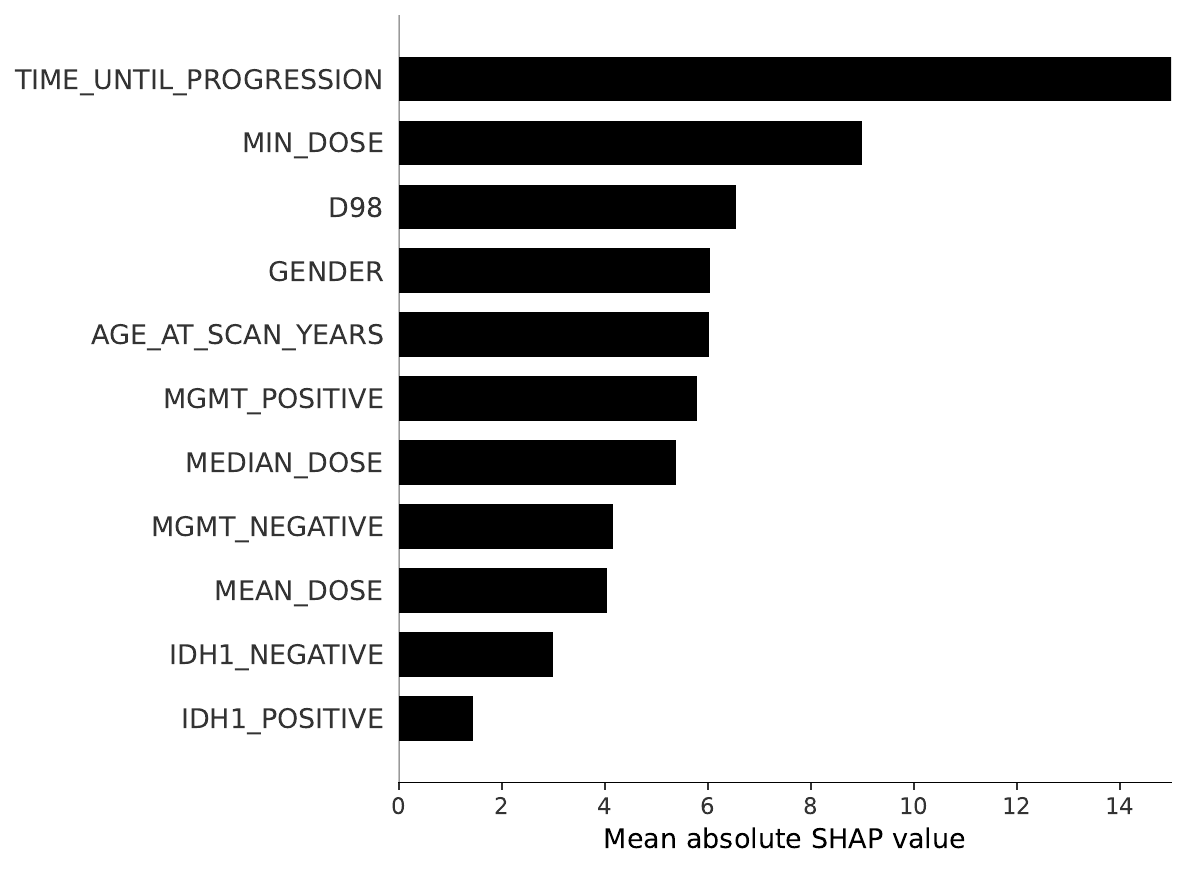}
    \caption{Mean feature importance analysis using SHAP in a 5-fold cross-validation setup.}
    \label{fig:SHAP}
\end{figure}

To understand the factors contributing most to the model's predictions, we conducted a feature importance analysis using SHAP analysis on the validation sets. The results illustrated in \Cref{fig:SHAP} show that the Isocitrate dehydrogenase (IDH) gene mutation status has the least impact on the model's outcome. On the other hand, the time between the last treatment and the first progression has the greatest influence on the model's prediction. Furthermore, out of the included dose-related features, the minimum dose, and the near-minimum dose (D98) of the enlarging lesion are the most relevant for progression status classification. 

\section*{Discussion}
The differentiation of pseudoprogression from true progression is essential for optimal clinical decision-making and patient management following chemoradiation for glioblastoma. The miss-classification of one over the other can result in unnecessary and potentially unfavorable alteration of therapeutic regimes. This involves starting new interventions that may be unnecessary or even hazardous due to side effects, or the premature termination of successful treatments. Moreover, distinguishing between these two events is essential for assessing the effectiveness of therapies in clinical trials, which impacts the precise evaluation of innovative glioblastoma therapy approaches. Novel deep learning techniques could address this critical challenge. 

In this work, we present a multimodal transformer-based deep learning model for progression classification in glioblastoma patients. Equipped with the self-supervised learning approach, the model can leverage the more abundant unlabeled brain MR scans to improve classification performance in data-scarce applications. Furthermore, the transformer's self-attention mechanism allows the model to learn the long-term dependencies that inherently exist within high-dimensional inputs, such as MRI data. Moreover, the proposed model utilizes guided cross-modal attention, allowing the model to benefit from the incorporation of different modalities. The proposed model achieves promising predictive performance with an AUC of 0.753, accuracy of 0.75, sensitivity of 0.727, and specificity of 0.8, outperforming the existing state-of-the-art methods (\Cref{tab:SOTA}). The improvement in predictive performance can be attributed to a multitude of factors. The proposed model uses the whole-brain MRI volume to make predictions, compared to a limited number of slices used in the other baselines. This allows the model to benefit from a more comprehensive spatial understanding of the tumor's characteristics. Furthermore, the proposed model employs guided cross-attention to integrate the information from the different input modalities, allowing the model to learn more refined discriminative features, in contrast to the direct concatenation approach used by the baseline models \cite{akbari2020histopathology, jang2018prediction, sun2021differentiation}.

% TODO: discuss the results in Table 3. 
This work investigated the predictive capability of each modality and the impact of integrating them into a unified model. As shown in \Cref{tab:resultsMain}, the clinical data outperformed the imaging modality across all metrics except for specificity. Moreover, the results also show that the model has successfully learned to integrate complementary features that are not available when using a single modality, outperforming the models of single modalities. 

To investigate the impact of the learning strategy of the MRI encoder on the predictive performance, we compare the performance of the proposed approach with transfer learning as well as end-to-end training. As shown in \Cref{tab:resultAblation}, the transfer learning approach performed $0.727$, $0.65$, $0.636$, and $0.7$ in terms of AUC, accuracy, sensitivity, and specificity, respectively. The transfer learning method outperformed the training from scratch approach, which achieved $0.697$, $0.6$, $0.545$, and $0.667$ for the AUC, accuracy, sensitivity, and specificity, respectively. The notable improvement in the performance provided by the self-supervised encoder model relative to the other methods can be attributed to the ability of the former to learn nuanced features and patterns from the unlabeled data in the pre-training process. This contrasts with the other approaches, which rely on labeled data and may struggle with overfitting or insufficient representation learning, especially when labeled data is limited \cite{Chen2019Self, shurrab2022self}.

The previous results indicate that the model benefits from incorporating the structured clinical features. However, in this context, not all features contribute equally to the deep learning progression classification, which is illustrated in \Cref{fig:SHAP}. This graph shows that the time between the treatment termination and the time point at which progression was observed has the greatest impact on the model's output compared to the other clinical features. This observation is aligned with the findings in the literature which states the PsP typically occurs within the first $3$ months after the end of the treatment for roughly $60\%$ of progression cases \cite{da2011pseudoprogression, brandsma2008clinical}. Moreover, among the dose-related features,  the minimum and the near-minimum dose (D98) of the enlarging lesion play a considerable role in determining the model's output. However, the other dose-related features have a lesser impact on the prediction outcome. The minimum dose and the D98 (near-minimum dose) of the enlarging lesion can indicate, whether the enlarging lesion is entirely located in the high-dose region of the previous radiation therapy. As the radiotherapy dose causes pseudoprogression but impedes tumor progression, a higher D98 and minimum dose is expected for pseudoprogression than for real tumor progression.
For the molecular-pathologic markers, MGMT methylation status has a greater impact on the model's prediction than IDH1. 
This is also in line with the available literature showing that pseudoprogression is more likely to occur in MGMT-methylated glioblastomas \cite{brandes2008mgmt}. Taken together, the feature importance analysis indicates that the model is able to successfully learn the most relevant clinical and dosimetric features for differentiating PsP from TP.

While this work provides insights into the task of progression classification, it is essential to mention the limitations to prevent overgeneralization of the reported results. First, as typical for pseudoprogression datasets, the sample size of the two datasets used in this study is relatively limited (n = 59 and n = 20), which might introduce biases and limit the generalizability of the findings. Second, the model utilizes only two MRI sequences, namely T1CE and FLAIR. The model could benefit greatly by including additional advanced MRI modalities such as diffusion-weighted imaging (DWI) and perfusion MRI (PWI) as well as metabolic MRI-CEST \cite{zhou2022review}, which could provide further insights into tissue characteristics and improve the accuracy and robustness of the model's predictions \cite{wen2010updated, henson2008brain, smith2005serial}. The third limitation is the relative lack of straightforward interpretability offered by deep learning models.

In conclusion, this work introduces a multimodal transformer-based deep learning model that improves the differentiation between pseudoprogression and true progression in glioblastoma patients in a non-invasive manner. Owing to the self-supervised pre-training, the model can achieve a competitive performance despite the limited training data. Furthermore, the guided cross-attention employed by this model demonstrates an effective approach to integrating information from imaging data, clinical, molecular-pathologic, and dosimetric parameters, resulting in improved predictive accuracy and outperforming the state-of-the-art methods.

\section*{Data availability}
The datasets generated during and/or analyzed during the current study are available from the corresponding author upon reasonable request.

% \bibliography{sample}
\bibliography{main}

\section*{Author contributions statement}
The study was designed and conceived by FP, YH, and AG. AG wrote the program code and conducted the experimental work. AS contributed to the code. AG, AS, and FP wrote the original draft. SS, FP, AG, CS, MAS, and BF collected and processed the data. MAS and AD provided the imaging data for this study. MAS and AD provided imaging expertise. CB, KB, RF, AD, OS, RC, DD, JS, AM, SB, SAS, DH, PH, UG and MAS reviewed and edited the script. All authors have read and approved the script.

% Must include all authors, identified by initials, for example:
% A.A. conceived the experiment(s),  A.A. and B.A. conducted the experiment(s), C.A. and D.A. analysed the results.  All authors reviewed the manuscript. 

\section*{Additional information}

\textbf{Competing interests:} The authors declare no competing interests.

% Figures and tables can be referenced in LaTeX using the ref command, e.g. Figure \ref{fig:aucProposed} and Table \ref{tab:resultsMain}.

\end{document}